\documentclass{mpe_report}

\usepackage{psfig,graphicx,epsfig}
\usepackage{color}
\usepackage{amsmath,amssymb,epic,eepic,array}

\begin{document}

%\pagenumbering{roman}
%
%\include{coverpage}  \clearpage
%\thispagestyle{empty}\phantom{x}\clearpage
%
%\include{coverpage2} \clearpage
%\thispagestyle{empty}\phantom{x}\clearpage
%
%\include{./Lesch/pref}  \clearpage
%\thispagestyle{empty}\phantom{x}\clearpage
%
%\include{participants}  \clearpage
%\thispagestyle{empty}\phantom{x}\clearpage
%
%\include{program} \clearpage
%
%
%\pagenumbering{Roman}
%\setcounter{page}{115}
%
%\include{tableofcontents} \clearpage
%

\pagenumbering{arabic}
\setcounter{page}{5}
 \renewcommand{\FirstPageOfPaper }{  5}\renewcommand{\LastPageOfPaper }{  8}

\title{Comptonization in the X-ray Spectra of \\ Radio Millisecond Pulsars}
\author{Slavko Bogdanov, Jonathan E. Grindlay, George B. Rybicki}  
\institute{Harvard-Smithsonian Center for Astrophysics, 60 Garden Street, Cambridge, MA 02138, U.S.A.}
\authorrunning{Bogdanov, Grindlay, \& Rybicki}

\maketitle

\begin{abstract}
The majority of X-ray-detected rotation-powered millisecond pulsars
(MSPs) appear to exhibit predominantly thermal emission, believed to
originate from the heated magnetic polar caps of the pulsar. In the
nearest MSP, J0437--4715 a faint PL is also observed at $>$3 keV,
usually associated with magnetospheric emission processes. However,
the hard emission in this and other similar MSPs may instead be due to
weak Comptonization of the thermal polar cap emission by energetic
electrons/positrons of small optical depth most likely in the pulsar
magnetosphere. This spectral model implies that all soft X-rays are of
purely thermal origin, which has important implications in the study
of neutron stars.
\end{abstract}

\section{Introduction}

With the advent of the current generation of X-ray telescopes,
\textit{Chandra} and \textit{XMM-Newton}, it has become apparent that
the population of rotation-powered millisecond pulsars (MSPs)
possesses very diverse spectral properties. The most energetic
($\dot{E}\sim10^{36}$ ergs s$^{-1}$), ``Crab-like'' MSPs, B1937+21,
B1821--24 (in the globular cluster M28), and J0218+4232, exhibit
strongly pulsed non-thermal X-rays with $L_X\sim10^{33}$ ergs
s$^{-1}$, which are attributed to non-thermal radiation processes in
the pulsar magnetosphere (Nicastro et al. 2004; Becker et al. 2003;
Rutledge et al. 2004; Webb, Olive, \& Barret 2004). Several binary
MSPs are observed at X-ray energies (with $L_X\sim 10^{31}$ ergs
s$^{-1}$) due to interaction of their energetic particle wind
($\dot{E}\sim10^{34-35}$ ergs s$^{-1}$) with matter from a close
stellar companion (Grindlay et al. 2002; Stappers et al. 2003;
Bogdanov, Grindlay, \& van den Berg 2005).  The bulk of X-ray detected
MSPs show predominantly thermal spectra with $L_X\sim10^{30-31}$ ergs
s$^{-1}$ (see Becker \& Tru\"mper 1999; Zavlin 2006; Bogdanov et
al. 2006; review talk by J. E. Grindlay). This thermal emission is
widely believed to arise due to heating of the magnetic polar caps by
a backflow of relativistic particles from the magnetosphere (see
e.g. Harding \& Muslimov 2002; Zhang \& Cheng 2003 and references
therein).

Thermal radiation from neutron stars (NSs) is of great interest in
astrophysics as it serves as a probe of the environment near the
surface of these compact objects.  In particular, the study of this
emission may shed light on the poorly understood properties of the
stellar interior, thus, providing constraints on the elusive NS
equation of state (EOS). Pavlov \& Zavlin (1997) have shown that X-ray
spectral and timing observations of MSPs can be used to measure
fundamental NS parameters such as the mass-to-radius ratio ($M/R$) of
the underlying NS. These objects are much better suited for such an
analysis than other NS systems such as X-ray binaries, isolated NSs,
and normal pulsars. In the latter objects there are numerous
complications arising due to the effect of the strong magnetic field
on the emergent radiation (Zavlin et al. 1995), the unknown
temperature distribution across the surface, severe reprocessing of
the thermal radiation by the magnetosphere, the uncertain altitude
above the NS surface (e.g. in X-ray bursts, see Cottam et al. 2002),
etc. Their low magnetic fields ($\sim$$10^{8-9}$ G), point-like
emission regions ($<$3 km), and ``clean'', non-variable emission make
MSPs potential laboratories for tests of fundamental NS physics.

However, it is important to realize that the thermal photons
originating at the surface of the NS must propagate through the
tenuous plasma present in the pulsar magnetosphere and wind before
reaching the observer.  Thus, it is essential to examine the effect of
this plasma on the thermal photons as it may impede efforts to
constrain the NS properties described above.

\begin{figure}
\centerline{\psfig{file=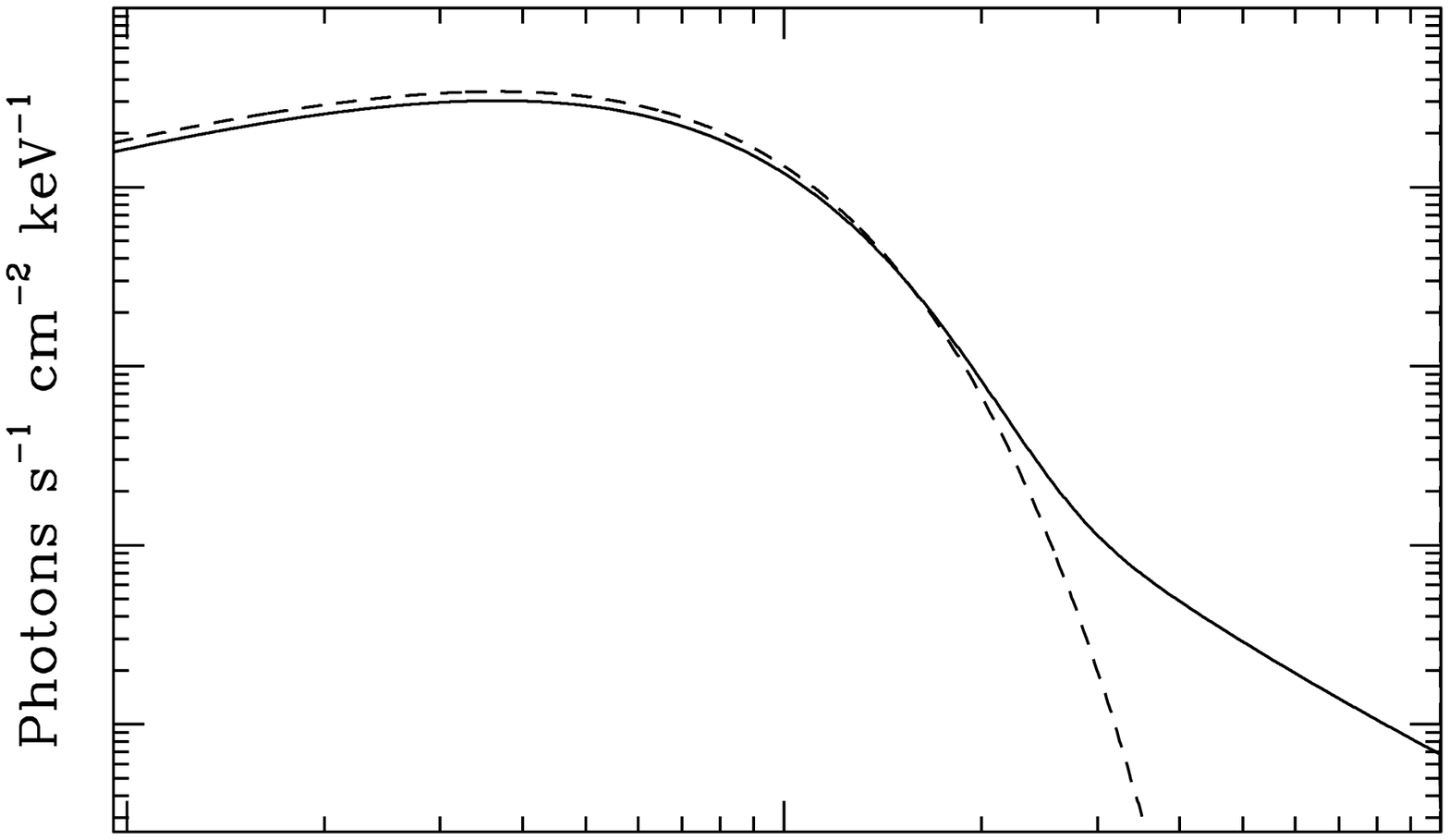,width=0.43\textwidth,clip=} }
\centerline{\psfig{file=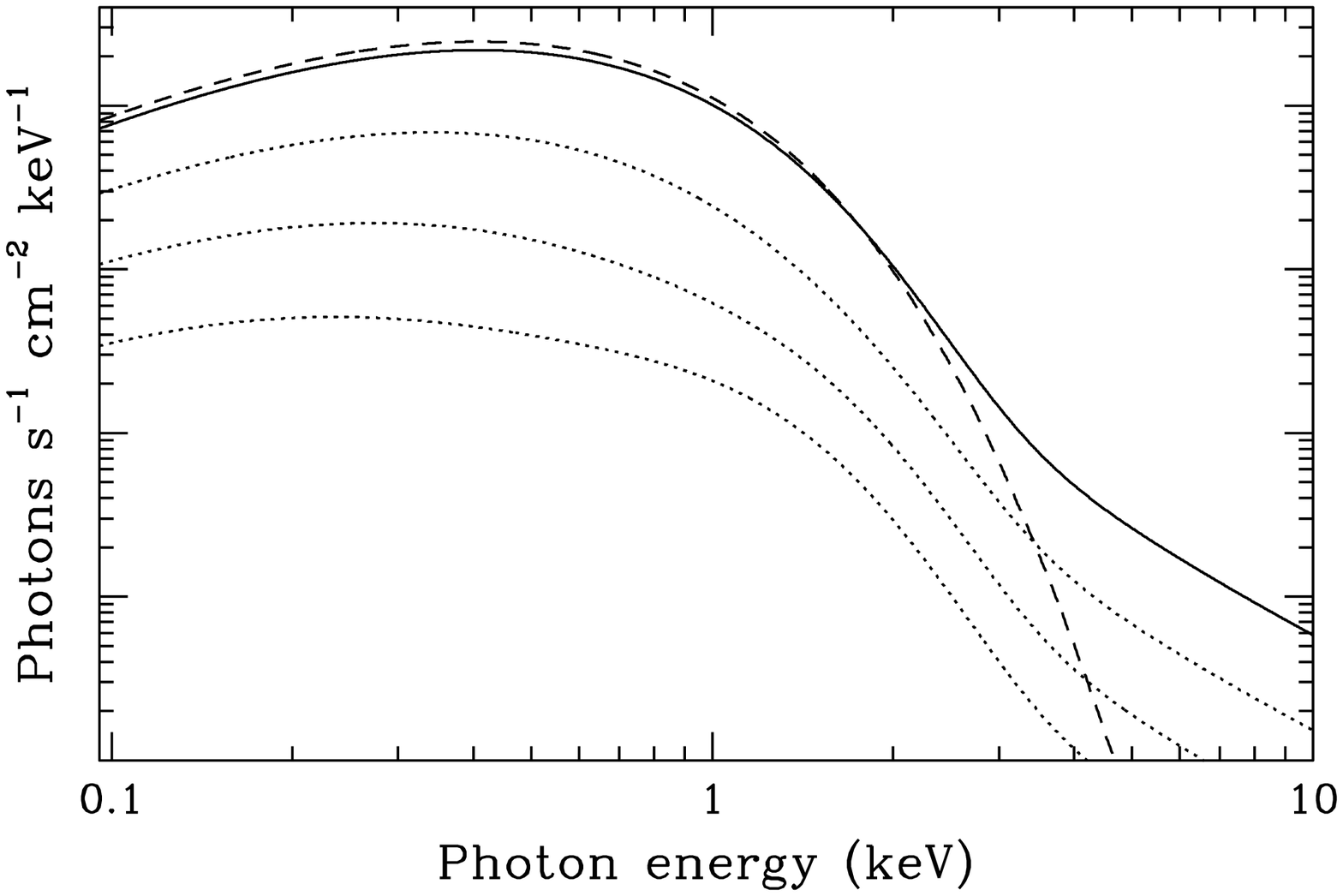,width=0.43\textwidth,clip=} }
\caption{(\textit{top}) Comptonized blackbody model spectrum for
$T_{\rm eff}=2.5$ MK, $kT_e=100$ keV, and $\tau=0.1$.
(\textit{bottom}) Comptonized H-atmosphere model spectrum for $T_{\rm
eff}=1.3$ MK, $kT_e=100$ keV, and $\tau=0.1$. The photon energies have
been corrected for the gravitational redshift assuming $R=10$ km and
$M=1.4$ M$_{\odot}$.  The dotted lines correspond to different viewing
angles relative to the surface normal ($\cos \theta = 0.5$, 0.25, and
0.1, from top to bottom, respectively), while the solid line is for
$\cos\theta=1$.  In both panels the dashed line represents the initial
(unscattered) thermal spectrum.
\label{image}}
\end{figure}

\begin{figure}[t!]
\centerline{\psfig{file=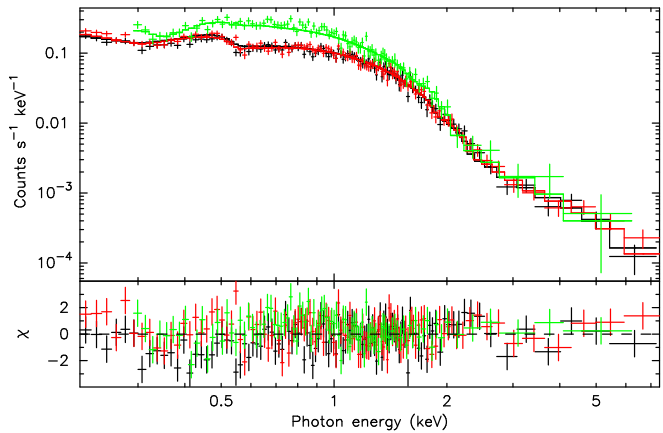,width=0.49\textwidth,clip=} }
\caption{\textit{Chandra} ACIS-S and
\textit{XMM-Newton} MOS1/2 spectra of PSR J0437--4715 fitted with a
two-temperature Comptonized thermal spectrum (see Bogdanov, Grindlay,
\& Rybicki 2006 for best fit parameters).
\label{image}}
\end{figure}

\begin{figure}[t!]
\centerline{\psfig{file=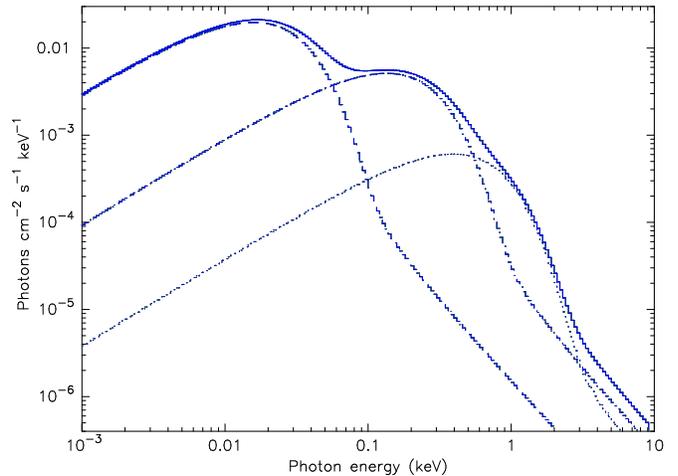,width=0.49\textwidth,clip=} }
\caption{Model for the broadband spectrum of PSR J0437-4715 as
inferred from X-ray and FUV data for this pulsar. The two hard thermal
components (observed in X-rays) originate from the heated magnetic
polar caps of the pulsar, while the softest ($\sim$$10^5$ K) component
(seen in the FUV) is due to emission from the rest of the NS surface
(see Kargaltsev, Pavlov, \& Romani 2004). Note the PL tail for each
thermal component due to Comptonization.
\label{image}}
\end{figure}

\section{The power-law tail}

The most suitable object for such a study is the nearest and brightest
radio millisecond pulsar known, PSR J0437--4715 (Johnston et al. 1993;
van Straten et al. 2001; Hotan, Bailes, \& Ord 2006).  This MSP was
the first to be detected at X-ray energies in the \textit{ROSAT}
all-sky survey (Becker \& Tr\"umper 1993). Subsequent \textit{ROSAT},
\textit{Chandra}, and \textit{XMM-Newton} observations (Zavlin \&
Pavlov 1998; Zavlin et al. 2002; Zavlin 2006) have revealed that the
0.1--10 keV emission consists of two thermal components, and a faint
power-law (PL) tail, with photon index $\Gamma\sim 2$. The latter can
only be clearly seen in the spectrum for $>$2.5 keV.  Due to the
limited photon statistics beyond $\sim$3 keV, the nature of this X-ray
component is ambiguous but is usually attributed to non-thermal
emission processes in the pulsar magnetosphere. However, for
J0437--4715 this model encounters difficulties when extrapolated to
lower energies where it is in disagreement with the FUV data for this
pulsar (Kargaltsev, Pavlov, \& Romani 2004) unless a break exists in
the PL below 0.1 keV.  Shock emission can also ruled out as the
strength of the pulsar wind ($\dot{E}=3.8\times10^{33}$ ergs s$^{-1}$)
is insufficient to have a significant effect on the He-WD companion. A
third thermal component is also rather implausible as it requires a
peculiarly small emission area ($\sim$ a few meters) and high surface
temperature ($T_{\rm eff}> 8\times 10^6$ K).
 
One physical mechanism for production of a PL spectrum in pulsars that
is often overlooked is inverse Compton scattering (ICS).  As a thermal
photon emitted at the NS surface propagates through the tenuous (low
optical depth) magnetospheric plasma and the particle wind of the
pulsar, it may be scattered by energetic e$^{\pm}$ and in the process
acquire energy.  In the low optical depth ($\tau<1$) regime,
\textit{repeated} ICS of thermal seed photons produces a PL
distribution of photons (see Rybicki \& Lightman 1979; Nishimura,
Mitsuda, \& Itoh 1986). Figure 1 illustrates the effect of
Comptonization on an initially thermal spectrum for both blackbody and
unmagnetized hydrogen atmosphere (Romani 1987; Zavlin, Pavlov, \&
Shibanov 1996; Heinke et al. 2006) spectra, using a model based on the
Comptonization algorithm of Nishimura et al. (1986).

As evident in Figure 2, the {\tt compbb} Comptonization model
available in
XSPEC\footnote{http://heasarc.gsfc.nasa.gov/docs/xanadu/xspec/} is
quite succesful at reproducing the observed shape of the X-ray
spectrum of J0437--4715.  Note that the {\tt compbb} model assumes a
thermal distribution of scattering particles.  However, in the pulsar
magnetosphere and the wind a thermal population of e$^{\pm}$ is not
sustainable due to the strong induced electric field. The scattering
e$^{\pm}$ are, therefore, most likely non-thermal (PL).  Nonetheless,
the model we have used remains valid since a PL photon spectrum is
produced for an arbitrary distribution of e$^{\pm}$ energies.  Note
that unlike the non-thermal emission PL, the Comptonization PL does
not extend into the FUV (see Fig. 3), thus, avoiding any conflict with
optical/UV data encountered by the alternative model.

\begin{figure}[t!]
\centerline{\psfig{file=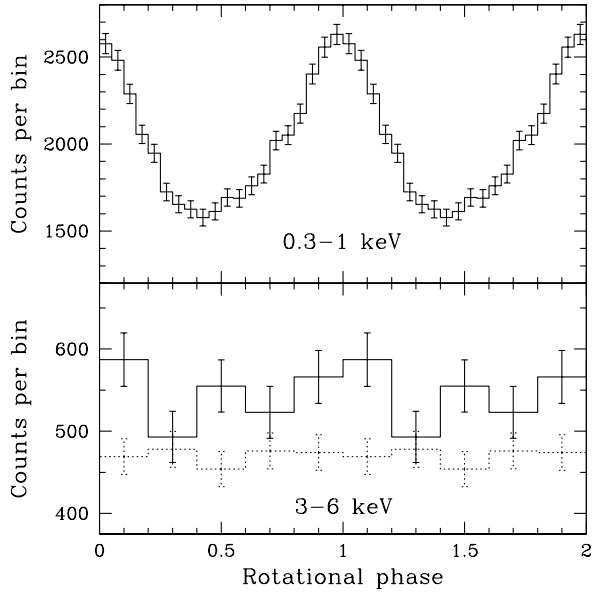,width=0.45\textwidth,clip=} }
\caption{\textit{XMM-Newton} EPIC-pn lightcurves of J0437 in the
0.3--1 keV (\textit{top}) and 3--6 keV (\textit{bottom}) bands. In the
Comptonization model these bands contain purely thermal (unscattered)
and Comptonized photons, respectively. The dotted line in the bottom
panel shows the background level.
\label{image}}
\end{figure}

Using a simple calculation we can get a sense of the density and
physical size of the scattering medium neccessary to have a
significant effect on the thermal spectrum. Assuming a
Goldreich-Julian (GJ) charged particle density in the MSP
magnetosphere, we obtain $n=n_{GJ}\sim0.07B/P = 4\times10^{9}$
cm$^{-3}$ (Goldreich \& Julian 1969), where we have assumed
$B\sim10^8$ G, and $P=5.76$ ms, aprropriate for J0437--4715. The
optical depth is given by $\tau_s=nL\sigma$, where $L$ is the path
length through the scattering medium and $\sigma$ is the scattering
cross section. For $L$ equal to the light cylinder radius
($r_{lc}=cP/2\pi=275$ km for J0437--4715) and $\sigma$ equal to the
Thomson cross section ($\sigma_T=0.24\times10^{-24}$ cm$^{2}$) we find
$\tau\sim7\times10^{-8}$, too low to have an observable effect on the
X-ray spectrum.  However, the actual particle density in the vicinity
of a pulsar may greatly exceed the GJ density\footnote{In the double
pulsar system PSR J0737--3039 (Burgay et al. 2003), there is
indication that the $e^{\pm}$ density in the wind of pulsar A may be
more than 4 orders of magnitude greater than predicted by pulsar
models (Lyutikov 2004; Arons et al. 2005).}, e.g., due to pair
production cascades by high energy photons.  Thus, $n\gg n_{GJ}$
and/or $L\gg r_{lc}$ are necessary to produce a sufficient scattering
optical depth to account for the PL tail in J0437--4715.

Insight into the geometry of the Comptonizing plasma may, in
principle, be obtained by observing the behavior of the PL radiation
as a function of the spin phase of the MSP. Figure 3 shows lightcurves
of J0437--4715 folded at the MSP spin period for the 0.3--1 keV and
3--6 keV energy ranges.  Unfortunately, the currently available timing
data do not permit any useful constraints on the properties of the
Comptonizing medium.  The future generation of X-ray facilities
(\textit{Constellation-X}\footnote{http://constellation.gsfc.nasa.gov/index.html}
and
\textit{XEUS}\footnote{http://www.rssd.esa.int/index.php?project=XEUS})
should be able to shed more light via phase-resolved spectroscopy at
energies beyond 10 keV.

\begin{figure}[t!]
\centerline{\psfig{file=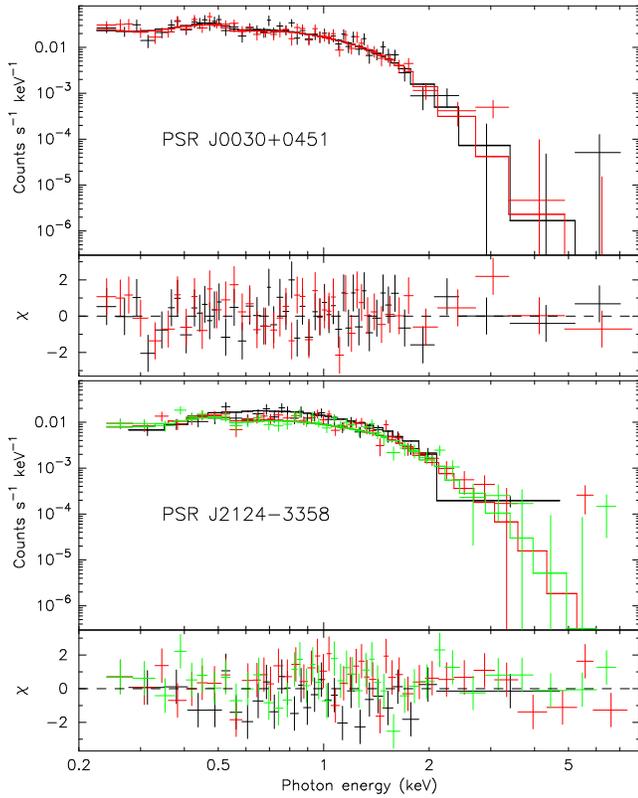,width=0.47\textwidth,clip=} }
\caption{(\textit{top}) \textit{XMM-Newton} MOS1/2 spectra of PSR
J0030+0451. (\textit{bottom}) \textit{Chandra} ACIS-S and
\textit{XMM-Newton} MOS1/2 spectra of PSR J2124-3358. In both cases
the data are fitted with a two-temperature blackbody spectrum. 
\label{image}}
\end{figure}

\section{Conclusion}
  
We have shown that the faint PL tail in the spectrum of the
millisecond pulsar PSR J0437--4715 could be the observable signature
of weak Comptonization of the thermal X-ray photons by particles in
the pulsar magnetosphere and wind.  Deep multiwavelength (X-ray and
optical/UV) observations of other MSPs, such as the nearby isolated
PSRs J0030+0451 and J2124--3358, would allow an important test of the
Comptonization model. For these MSPs, due to the limited count
statistics in the currently available data, the presence of a PL tail
cannot be established (see Fig. 4).  The absence of a stellar
companion for J0030+0451 and J2124--3358 has allowed very deep optical
observations, which have found no emission down to $V \sim 27-28$
(Koptsevich et al. 2003; Mignani \& Becker 2003).  Such observations,
coupled with more detailed models describing the spatial and energy
distribution of the particles populating the pulsar magnetosphere and
wind will lead to a more complete picture of the pulsar environment
and reveal the true nature of the PL emission from J0437--4715.

If Comptonization is indeed the mechanism responsible for the PL
emission, X-ray spectra of MSPs can potentially serve as a valuable
diagnostic of the charged particle population near the pulsar, thus,
providing important constraints for pulsar models. Perhaps more
importantly, the Comptonization model implies that \textit{all} soft
emission ($<$1 keV) from J0437--4715 is purely thermal (unscattered)
radiation, which has profound applications in the study of NS
structure.

\begin{acknowledgements}
The work presented here was supported in part by NASA grant AR6-7010X.
\end{acknowledgements}

        \clearpage

\end{document}